\documentstyle[multicol,aps,psfig]{revtex}

\begin{document}
\preprint{LBNL-42545}
%

\title{Systematic Study of High $p_T$ Hadron Spectra
  in pp, pA and AA Collisions from SPS to RHIC Energies}
\author{Xin-Nian Wang}
\address{Nuclear Science Division, Mailstop 70A-3307,\\
Lawrence Berkeley National Laboratory, Berkeley, CA 94720 USA}
\date{November 17, 1998}

\maketitle

\begin{abstract}

High-$p_T$ particle spectra in $p+p$ ($\bar p + p$), $p+A$ and $A+B$
collisions are calculated within a QCD
parton model in which intrinsic transverse momentum, its broadening due to
initial multiple parton scattering, and jet quenching due to parton energy
loss inside a dense medium are included phenomenologically. The intrinsic 
$k_T$ and its broadening in $p+A$ and $A+B$ collisions due to initial
multiple parton scattering are found to be very important at low energies (
$\sqrt{s}<50$ GeV). Comparisons with $S+S$, $S+Au$ and
$Pb+Pb$ data with different centrality cuts show that the differential 
cross sections of large transverse momentum pion production ($p_T>1$ GeV/$c$)
in $A+B$ collisions scale very well with the number of
binary nucleon-nucleon collisions (modulo effects of multiple initial
scattering). This indicates that semi-hard parton scattering is the dominant
particle production mechanism underlying the hadron spectra at moderate $p_T
\stackrel{>}{\sim} 1$ GeV/$c$. However, there is no evidence of jet
quenching or parton energy loss. Within the parton model, one can exclude an
effective parton energy loss $dE_q/dx>0.01$ GeV/fm and a mean free path
$\lambda_q< 7$ fm from the experimental data of $A+B$ collisions at the SPS
energies. Predictions for high $p_T$ particle spectra in $p+A$ and $A+A$
collisions with and without jet quenching at the RHIC energy are also given. 
Uncertainties due to initial multiple scattering and nuclear
shadowing of parton distributions are also discussed.

\end{abstract}


\begin{multicols}{2}

\section{Introduction}

Large-$E_T$ partons or jets are good probes of the dense matter formed
in ultra-relativistic heavy-ion collisions \cite{GP,WG92}, since they are
produced in the earliest stage of heavy-ion collisions and their
production rates are calculable in perturbative QCD. If a dense
partonic matter is formed during the initial stage of a heavy-ion
collision with a large volume and a long life time
(relative to the confinement scale $1/\Lambda_{\rm QCD}$), the produced large
$E_T$ parton will interact with this dense medium and, according to many
recent theoretical studies \cite{GW1,BDPS,BDMPS}, will lose its energy
via induced radiation. The energy loss is shown to depend on the
parton density of the medium. Therefore, the study of parton
energy loss can shed light on the properties of the dense matter in
the early stage of heavy-ion collisions. It is also a crucial test
whether there is any thermalization going on in the initial stage of
heavy-ion collisions.

Even though one cannot measure directly the
energy loss suffered by a high energy parton propagating through a
dense medium, its effective fragmentation functions must change and
leading particles must be suppressed due to the parton energy
loss. This can be measured directly in deeply inelastic $e+A$
collisions (for jets going through a normal nuclear matter)
\cite{GP2,WH} and direct-photon-tagged jets in high-energy heavy-ion 
collisions \cite{WH,WHS}. Since the large $p_T$ single-inclusive
particle spectra in nuclear collisions are a direct consequence of jet
fragmentation as in hadron-hadron and hadron-nucleus collisions, they
are also shown \cite{wang98} to be sensitive to parton energy loss or 
modification of jet fragmentation functions inside a dense medium. 

Because of the extremely small cross sections of large $p_T$ particle
production at the CERN SPS energy ($\sqrt{s}\approx 20$ GeV), measurements
of particle spectra in  heavy-ion collisions at moderately large $p_T$ have
not become available before recent experiments \cite{wa80,wa98}. In a recent 
paper \cite{wang98-2}, the author used a QCD parton model calculations to 
analyze the large $p_T$ spectra and found that there is no evidence of
parton energy loss in high-energy heavy-ion collisions. From the analysis of
the spectra, one can limit the effective energy loss to less than
$dE/dx<0.01 GeV/$fm. Since experimental data have shown evidences of an
interacting hadronic matter in heavy-ion collisions at the CERN SPS energy,
one can also conclude that the hadronic matter dose not cause apparent jet
energy loss.

In this paper, we will conduct a systematic study of high $p_T$ hadron
spectra in $pp$, $pA$ and $AB$ collisions, since the reliability of the
determination of a small effective parton energy loss in high-energy
heavy-ion collisions crucially depends on
the precision to which we understand the spectra without parton
energy loss. We will include both the intrinsic $k_T$ in $p+p$ collisions
and $k_T$ broadening due to multiple initial-state scattering in
$p+A$ and $A+A$ collisions in a phenomenological manner, 
which we find necessary to describe large $p_T$ particle
production at energies below $\sqrt{s}<50$ GeV. We will then compare the
calculations with the experimental data for $S+S$, $S+Au$ and $Pb+Pb$
collisions at the CERN SPS with different centrality cuts and verify the
scaling behavior characteristic of hard processes. Using the same parton
model we will then predict the high $p_T$ spectra in $pA$ collisions at the
RHIC energy and then calculate same spectra in $AA$ collisions with and
without effects of parton energy loss. The goal of this paper is to provide a
practical and phenomenological description of the high $p_T$ hadron spectra in
$p+p$ and $p+A$ collisions. This will provide a baseline prediction for
high-energy $A+A$ collisions against which we can study the effects of the
parton energy loss caused by the dense medium.

\section{$p+p$ Collisions: Initial $k_T$}

Large $p_T$ particle production in high-energy hadron-hadron
collisions has been shown to be a good test of the QCD parton model
\cite{owens}. Assuming partons inside the colliding
hadrons only have longitudinal momenta, the large $p_T$ particle production
cross section can be calculated as a convolution of elementary
parton-parton scattering cross sections, parton distributions inside the
hadron and parton fragmentation functions. The factorization theorem
ensures that the parton distributions and fragmentation functions are
universal and can be measured in other hard processes, {\it e.g.},
$e^+e^-$ annihilation, deeply inelastic $ep$ collisions. With high precision
data, parton distributions and fragmentation functions have been
parameterized including their QCD evolution with the momentum scale $Q$
of the hard processes \cite{mrs,bkk} which in turn can be used
to calculate particle production in many other hard processes. This
parton model with collinear approximation has been rather successful so far in
describing large $p_T$ 
particle production in high-energy $p\bar p$ collisions
\cite{bk95} with $\sqrt{s}>50$ GeV. 


    The above described collinear parton model fails to account the
experimental data on angular correlation of produced heavy quarks and the
total transverse momentum distribution of the heavy quark pairs \cite{charm1}
or the Drell-Yan lepton pairs \cite{dy}. In terms of parton models one
naturally expects that partons 
inside a hadron carry at least an average intrinsic transverse momentum of
about a few hundred MeV, reflecting the hadron size via the
uncertainty principle. Furthermore, higher order
pQCD processes like initial state radiation or $2\rightarrow 3$
subprocesses with additional radiated gluons can also give rise to
large initial transverse momentum for the colliding partons.
However, even within next leading order (NLO) pQCD collinear parton model,
experimental data of Dell-Yan lepton pairs production \cite{dy}, heavy quark
pair production \cite{charm2}, and direct photon production \cite{photon} all
indicate that an average initial parton transverse momentum of about 1 GeV is
needed to describe the data. Such a large value is an indication of its
perturbative nature. Theoretical techniques have been developed to include all
high order contributions\cite{altarelli,collins,ddd}, though it is still
problematic to differentiate what is true intrinsic and what is pQCD generated
initial transverse momentum. One can re-sum all the radiative corrections into
a ``Sudakov'' form factor up to a scale $Q_0$ and define
(non-perturbative) contributions below this scale as ``true'' intrinsic, even
though such an ``intrinsic'' average transverse momentum still depend on
colliding energy $\sqrt{s}$ and the momentum scale $Q$ of the hard
processes. Such schemes have been developed for processes, such as Drell-Yan
\cite{altarelli,collins,ddd} and heavy quark production \cite{charm3} where two
scales are involved, with only minimum phenomenological success in limited
kinematic region. The problem in processes such as direct photon and single
inclusive hadron production, where there is only a single scale, is even more
difficult \cite{sterman}.

In this paper, we will adopt a more phenomenological approach by introducing
initial parton transverse momentum with a Gaussian form of distribution as
has been done in Drell-Yan \cite{field}, heavy quark \cite{charm2}, direct
photon \cite{photon} and single inclusive hadron production
\cite{photon,feynman,owen2}. Since it is difficult to extend the NLO
calculation to $p+A$ and $A+A$ collisions where multiple collisions occur and
our main purpose is to provide an effective description of single inclusive
hadron spectra not only in $p+p$ but also in $p+A$ and $A+A$ collisions, we
will use lowest order (LO) parton model with a simple scheme of
including the intrinsic transverse momentum with a Gaussian distribution.
The inclusion of initial transverse momentum  can be considered as
a parameterization of both higher order and nonperturbative corrections. The
predictive power of this model then lies in the energy and flavor dependence of
the hadron spectra. Similar comments can be made for our modeling of multiple
parton scattering effect in $p+A$ and $A+A$ collisions.

    One can define parton distributions in both fractional 
longitudinal momentum $x$ and initial transverse momentum $k_T$ in a
factorized form,
\begin{equation}
dxd^2k_T g_N(k_T,Q^2) f_{a/N}(x,Q^2),
\end{equation}
where $g_N(k_T,Q^2)$ is the transverse momentum distribution and
$f_{a/N}(x,Q^2)$ are the normal parton distribution functions for which
we will use the MRS D-' parameterization \cite{mrs}.
The inclusive particle production cross section in $pp$ collisions
will then be given by \cite{owens}
\begin{eqnarray}
  \frac{d\sigma^h_{pp}}{dyd^2p_T}&=&K\sum_{abcd}
  \int dx_a dx_b d^2k_{aT} d^2k_{bT} g_p(k_{aT},Q^2) \nonumber \\
  &\times& g_p(k_{bT},Q^2)f_{a/p}(x_a,Q^2)f_{b/p}(x_b,Q^2) \nonumber \\
  &\times& \frac{D^0_{h/c}(z_c,Q^2)}{\pi z_c}
  \frac{d\sigma}{d\hat{t}}(ab\rightarrow cd), \label{eq:nch_pp}
\end{eqnarray}
where $D^0_{h/c}(z_c,Q^2)$ is the fragmentation function of parton $c$
into hadron $h$ as parameterized in \cite{bkk,bkk2} from $e^+e^-$ data,
$z_c$ is the momentum fraction of a parton jet carried by
a produced hadron. The $K\approx 2$ (unless otherwise specified) factor
is used to account for higher order QCD corrections to the jet production 
cross section \cite{xwke}.

    We define the momentum fraction $x$ in terms of the
light-cone variables $x_a=(E_a+p_{\parallel a})/\sqrt{s}$ for partons
in the forward beam direction and $x_b=(E_b-p_{\parallel b})/\sqrt{s}$
in the backward beam direction. The four-vector momenta for the
colliding partons are then $p_a=(E_a, {\bf k}_{Ta},p_{\parallel a})$,
$p_b=(E_b,{\bf k}_{Tb},p_{\parallel b})$ with
\begin{eqnarray}
E_a&=&(x_a\sqrt{s}+\frac{k^2_{Ta}}{x_a\sqrt{s}})/2, \nonumber\\
p_{\parallel a}&=&(x_a\sqrt{s}-\frac{k^2_{Ta}}{x_a\sqrt{s}})/2, \nonumber \\
E_b&=&(x_b\sqrt{s}+\frac{k^2_{Tb}}{x_b\sqrt{s}})/2, \nonumber \\
p_{\parallel b}&=&(-x_b\sqrt{s}+\frac{k^2_{Tb}}{x_b\sqrt{s}})/2.
    \label{eq:mom4}
\end{eqnarray}

In principle, one should also include the transverse momentum smearing
from the jet fragmentation. We neglect this in our calculation of
particle spectra in the central rapidity region and consider it
been effectively included in the calculation by adjusting
the initial $k_T$ distribution. In this case, particles are produced in
the same direction of the fragmenting jet. The Mandelstam variables
for the elementary parton-parton scattering processes are then,
\begin{eqnarray}
\hat{s}&=&x_a x_b\sqrt{s} +\frac{k_{Ta}^2 k_{Tb}^2}{x_a x_b\sqrt{s}}
-2k_{Ta}k_{Tb}\cos(\phi_a-\phi_b), \nonumber \\
\hat{t}&=&-\frac{p_T}{z_c}\left( x_a\sqrt{s} e^{-y}
+\frac{k_{Ta}^2}{x_a\sqrt{s}}e^y - 2k_{Ta}\cos\phi_a\right), \nonumber \\
\hat{u}&=&-\frac{p_T}{z_c}\left( x_b\sqrt{s} e^y
+\frac{k_{Tb}^2}{x_a\sqrt{s}}e^{-y} - 2k_{Tb}\cos\phi_b\right),
\end{eqnarray}
where $p_T$ and $y$ are the transverse momentum and rapidity of the
produced particle, $\cos\phi_a={\bf k}_{Ta}\cdot {\bf p}_T/k_{Ta}p_T$,
and $\cos\phi_b={\bf k}_{Tb}\cdot {\bf p}_T/k_{Tb}p_T$. The momentum
fraction of the produced hadron $z_c$ is then given by the identity
for massless two-body scattering, $\hat{s}+\hat{t}+\hat{u}=0$.

In Eq.~(\ref{eq:nch_pp}), one should also restrict the initial
transverse momentum $k_{Ta}< x_a\sqrt{s}$ and
$k_{Tb}<x_b\sqrt{s}$ such that the partons' longitudinal
momenta in Eq.~(\ref{eq:mom4}) have the same signs as their parent
hadrons. In addition, one of the Mandelstam variables can approach to
zero if the initial $k_T$ is too large and then the parton cross
section could diverge. To avoid these problems, we introduce a
regulator $\mu^2$ in the denominators of the parton-parton cross
sections $d\sigma/d\hat{t}(ab\rightarrow cd)$. We choose $\mu=0.8$ GeV
in our following calculations. The resultant spectra are sensitive to
the choice of $\mu$ only at $p_T$ around $\mu$ where we believe that
QCD parton model calculation becomes unreliable.

We assume the initial $k_T$ distribution $g_N(k_T)$ to have a Gaussian form,
\begin{equation}
    g_N(k_T,Q^2)=\frac{1}{\pi \langle k^2_T\rangle_N} 
    e^{-k^2_T/\langle k^2_T\rangle_N}.
\end{equation}
Since our initial $k_T$ includes both the intrinsic and pQCD
radiation-generated transverse momentum, the variance in the Gaussian
distribution $\langle k^2_T\rangle_N$ should depend on the momentum
scale $Q$ of the hard processes. As pointed out by Owen and Kimel \cite{owen2},
in addition to the leading log contribution from initial-state radiation that
is included in the $Q^2$ dependence of the parton distributions, the recoil
effect of large-angle gluon emissions can lead to $Q$-dependent 
$\langle k^2_T\rangle_N$ in this approach of QCD parton model.
In the following we shall consider both a
constant $\langle k^2_T\rangle_N$ and one that depends on the momentum scale of
the hard processes. For the $Q$-dependent case, we choose the following 
form \cite{owen2} in our model
\begin{equation}
\langle k^2_T\rangle_N(Q^2)= 1.2 ({\rm GeV}^2) + 0.2\alpha_s(Q^2) Q^2.
\label{kperp}
\end{equation}
The form of the $Q$-dependence and the parameters are chosen to reproduce the
experimental data. Following the same approach as in 
Refs. \cite{feynman,owen2}, we choose $Q^2$ to 
be $Q^2=2\hat{s}\hat{t}\hat{u}/(\hat{s}^2+\hat{t}^2+\hat{u}^2)$.

\begin{figure}
\centerline{\psfig{figure=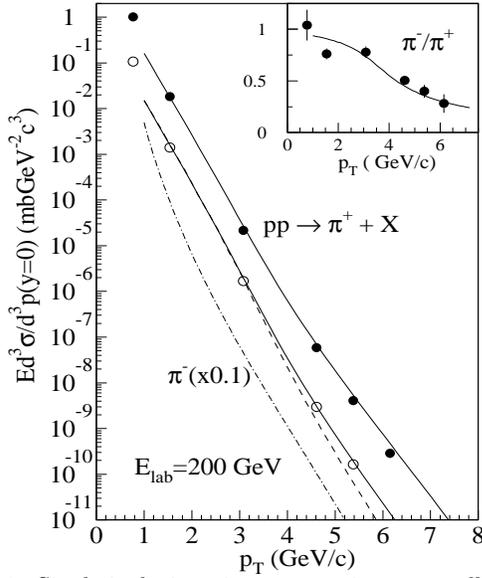,width=2.5 in,height=3.0in}}
\caption{ Single-inclusive pion spectra in $p+p$ collisions at 
$\protect E_{\rm lab}=200$ GeV. The solid lines are QCD parton model
calculations with $Q$-dependent intrinsic $k_T$ and the dot-dashed line is
without. The dashed line is for a constant intrinsic 
$\langle k^2_T\rangle_N=1.5$ GeV$^2$.  Experimental data are from
Ref.~\protect\cite{cronin-ex1}. The 
inserted figure shows the corresponding  $\protect\pi^-/\pi^+$ ratio.}
\label{figsps1}
\end{figure}

\begin{figure}
\centerline{\psfig{figure=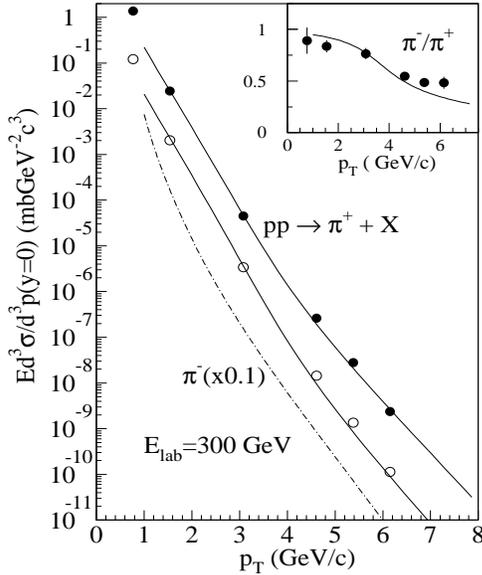,width=2.5 in,height=3.0in}}
\caption{ Single-inclusive pion spectra in $p+p$ collisions at 
$\protect E_{\rm lab}=300$ GeV. The solid lines are QCD parton model
calculations with $Q$-dependent intrinsic $k_T$ and the dot-dashed line is
without. Experimental data are from Ref.~\protect\cite{cronin-ex1}. The 
inserted figure shows the corresponding  $\protect\pi^-/\pi^+$ ratio.} 
\label{figsps2}
\end{figure}

\begin{figure}
\centerline{\psfig{figure=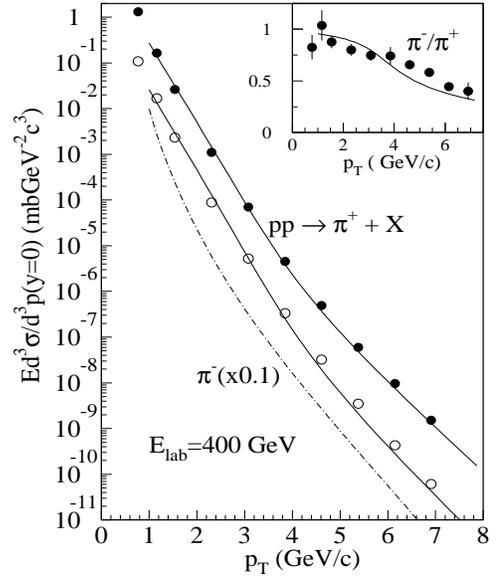,width=2.5 in,height=3.0in}}
\caption{ The same as Fig.~\protect\ref{figsps2}, except at 
  $\protect E_{\rm lab}=400$ GeV} 
\label{figsps3}
\end{figure}

\begin{figure}
\centerline{\psfig{figure=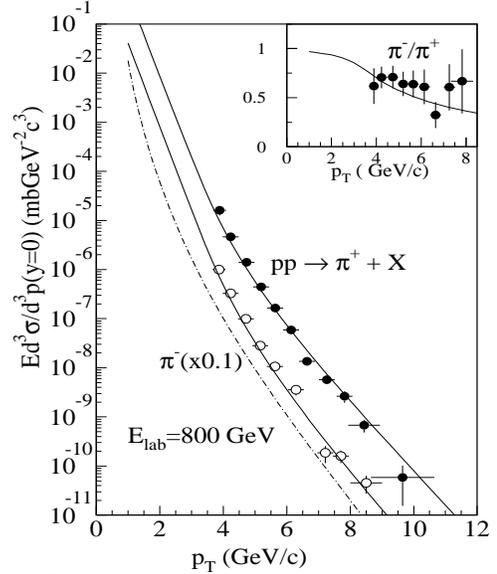,width=2.5 in,height=3.0in}}
\caption{ The same as Fig.~\protect\ref{figsps2}, except at 
  $\protect E_{\rm lab}=800$ GeV and the experimental data are from
   Ref.~\protect\cite{ppex2}} 
\label{figsps4}
\end{figure}

\begin{figure}
\centerline{\psfig{figure=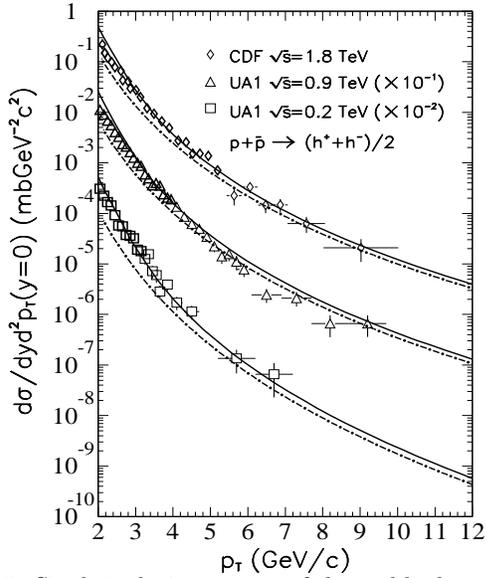,width=2.5 in,height=3.0in}}
\caption{ Single-inclusive spectra of charged hadrons in $p+\bar{p}$
  collisions at  $\sqrt{s}=200, 900, 1800$ GeV. The solid lines are QCD
  parton model calculations with intrinsic $k_T$ and the dot-dashed lines
  are without. Experimental data are from Refs.~\protect\cite{ua1pt,cdfpt}.}
\label{figsps5}
\end{figure}

    Shown in Figs.~\ref{figsps1}-\ref{figsps4} are our calculated
spectra for charged pions as compared to the experimental data
\cite{cronin-ex1,ppex2} for $p+p$ collisions at $E_{\rm lab}=$200,
300, 400, 800 GeV. 
As one can see from the figures that QCD parton model
calculations with the initial $k_T$
smearing (solid lines) as given in Eq.~\ref{kperp} fit the 
experimental data very well over all energy range. However, without 
the initial $k_T$ smearing (dot-dashed lines for $\pi^-$) the calculations 
significantly underestimate the experimental data, as much as a factor
of 20 at $E_{\rm lab}=200$ GeV. This is because the QCD
spectra are very steep at low energies and even a small amount of
initial $k_T$ could make a big increase to the final spectra. As the
energy increases, the QCD spectra become flatter and small amount of
initial $k_T$ does not change the spectra much, as we can already see by
comparing the spectra for $E_{\rm lab}=800$ GeV (Fig.~\ref{figsps4}) to that
of 200 GeV (Fig.~\ref{figsps1}). This is further demonstrated in
Fig.~\ref{figsps5} where we compare the LO calculation to experimental
data in high-energy $p\bar{p}$ collisions. Here we have used a $K=1.5$
factor. One can see that at higher collider energies, the initial $k_T$ does
not make much difference to the spectra at high transverse momentum. In the
case of no intrinsic $k_T$, we didn't use the 
regulator $\mu$ in our calculation. Instead, we used the $p_T$ of
particle as a cut-off of the phase integral in Eq.~(\ref{eq:nch_pp}).
This is why the spectra (dot-dashed lines) increase faster at smaller
$p_T$ than the ones with initial $k_T$ smearing in which a regulator
of $\mu=0.8$ GeV is used.

In Fig.~\ref{figsps1}, we also show as a dashed line the calculated spectrum of
$\pi^-$ with a constant average initial transverse momentum, 
$\langle k^2_T\rangle_N=1.5$ GeV$^2$. At large $p_T$ such a constant intrinsic
$\langle k^2_T\rangle_N$ underestimates the experimental data. This
demonstrates phenomenologically why the $Q$ dependence of 
$\langle k^2_T\rangle_N$ is needed to fit the experiment-


\noindent al data in particular
at large values of $x_T=2p_T/\sqrt{s}$ for the choice of $Q^2$ we used.
One can also use an alternative choice for $Q^2={P^{\rm jet}_T}^2=(p_T/z_c)^2$.
In this case we found that one does not need $Q$-dependent 
$\langle k^2_T\rangle_N$ to fit the data in $pp$ collisions. However, one
needs to use an energy-dependent initial parton transverse momentum 
distribution.
We also checked that as energy
increases such a $Q$-dependent $\langle k^2_T\rangle_N$ is not needed anymore
to fit the experimental data. However, the available experimental data at
collider energies are only limited to finite $p_T$ range where
$x_T=2p_T/\sqrt{s}$ is not so large as compared to the low energy data. In the
following we will use the $Q$-dependent initial momentum distribution 
\begin{figure}
\centerline{\psfig{figure=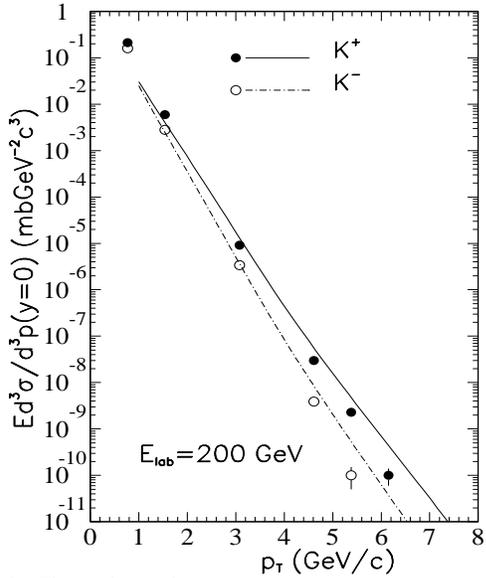,width=2.5 in,height=3.0in}}
\caption{ Single-inclusive spectra of charged kaons in $p+p$
  collisions at $\protect E_{\rm lab}=200$ GeV. The lines are QCD parton
  model calculations with intrinsic $k_T$ and data are from 
  Ref.~\protect\cite{cronin-ex1}.}
\label{figsps6}
\end{figure}

\noindent for our study of high $p_T$ spectra in $p+A$ and $A+A$ collisions.

\begin{figure}
\centerline{\psfig{figure=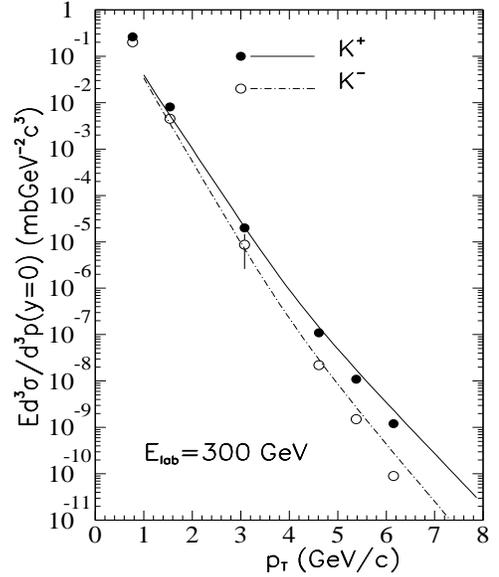,width=2.5 in,height=3.0in}}
\caption{ The same as Fig.~\protect\ref{figsps6}, except at 
  $\protect E_{\rm lab}=300$ GeV} 
\label{figsps7}
\end{figure}

In the inserted boxes in Figs.~\ref{figsps1}-\ref{figsps4}, we also plot
$\pi^-/\pi^+$ ratio as 
a function of $p_T$. At higher $p_T$, particle production is more dominated
by the leading hadrons from valence quark scattering. Since there are
more up-quarks than down-quarks in $p+p$ system, one should expect the
ratio to become smaller than 1 and decrease with $p_T$. The QCD parton model
calculations describe this isospin dependence of the spectra very
well. For other flavors of hadrons, we show in
Figs.~\ref{figsps5}-\ref{figsps7} the calculated kaon spectra at
different energies. We see that the agreement with experimental data
is very good, except at large $p_T$ at low energies which may be
improved by using better fragmentation functions for kaons. 
One can also notice that $K^-$ spectra at large $p_T$
are about a factor of 10 smaller than $K^+$. This is because the content of
strange quarks in a nucleon is much smaller than up (down) quarks which are
responsible for leading $K^+$ ($K^0$) hadron production. We did not
calculate and compare the spectra of produced protons and anti-protons
because there is no very accurate parameterization of the corresponding
fragmentation functions.

\begin{figure}
\centerline{\psfig{figure=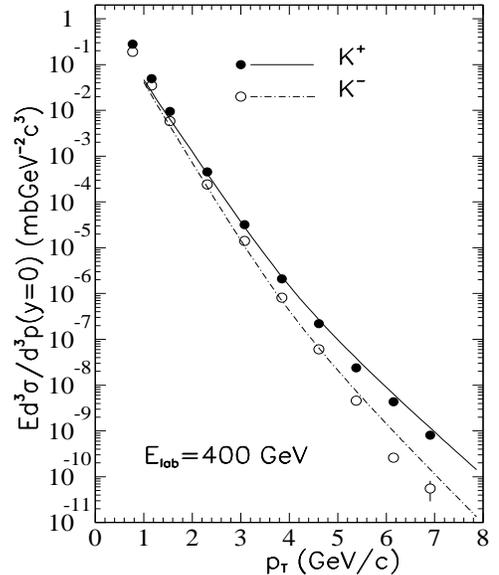,width=2.5 in,height=3.0in}}
\caption{ The same as Fig.~\protect\ref{figsps6}, except at 
  $\protect E_{\rm lab}=400$ GeV} 
\label{figsps8}
\end{figure}

\section{$p+A$ Collisions: $k_T$ broadening}

Since single-inclusive particle spectra at high $p_T$ in $p+A$
collisions have been shown, both experimentally \cite{cronin-ex1} and 
theoretically \cite{cronin-th1}, to be sensitive to multiple
initial-state scattering, or Cronin effect, it is important that
we take into account this effect here in order to have a quantitative
study of the change of particle spectra at high $p_T$ in
nucleus-nucleus collisions. One can study the Cronin effect in a
model of multiple parton scattering \cite{cronin-th1}. In such a
model, one uses Glauber multiple scattering formula to treat parton
scattering between beam and target partons which is essential to
incorporate the interference effect. For example, when contribution
from double scattering is considered, one must also include the
absorptive part of the single scattering which has a negative
contribution. This absorptive part then cancels part of the
contribution from double scattering. Consequently, the enhancement
of particle spectra because of double scattering decreases with $p_T$
in the form of $1/p_T^2$ and in general with $\sqrt{s}$
\cite{wang-rep}. One can also find that the dominant
contribution in double scattering comes from the case where one of
the scatterings is soft and large part of the final $p_T$ of the
produced jet comes from just one hard scattering. 
At much lower $p_T$ the absorptive correction is
much larger than the double scattering contribution and the spectra
there is even suppressed. So the integrated cross section or average
particle multiplicity at moderate $p_T$ are not affected by multiple
parton scattering even though the spectra are modified, as observed in the
DY case \cite{pa-DY}. This will provide us 
a justification for our following phenomenological treatment of
multiple parton scattering in $p+A$ and $A+A$ collisions.

    In this paper we assume that the inclusive differential cross
section for large $p_T$ particle production is still given by a single
hard parton-parton scattering. However, due to multiple parton
scattering prior to the hard processes, we consider the initial
transverse momentum $k_T$ of the beam partons is broadened. Assuming
that each scattering provide a $k_T$ kick which also has a Gaussian
distribution, we can in effect just change the width of the initial
$k_T$ distribution. Then the single inclusive particle cross section
in minimum-biased $p+A$ collisions is,
\begin{eqnarray}
  \frac{d\sigma^h_{pA}}{dyd^2p_T}=&K&\sum_{abcd} \int d^2b t_A(b)
  \int dx_a dx_b d^2k_{aT} d^2k_{bT} \nonumber \\  
  &\times& g_A(k_{aT},Q^2,b) f_{a/p}(x_a,Q^2) \nonumber \\
  &\times& g_p(k_{bT},Q^2) f_{b/A}(x_b,Q^2,b) \nonumber \\
  &\times& \frac{D^0_{h/c}(z_c,Q^2)}{\pi z_c}
  \frac{d\sigma}{d\hat{t}}(ab\rightarrow cd), \label{eq:nch_pA}
\end{eqnarray} 
where $t_A(b)$ is the nuclear thickness function normalized to $\int
d^2b t_A(b)=A$.  We will use Wood-Saxon form of nuclear distribution
for $t_A(b)$ throughout this paper unless specified otherwise. The parton
distribution per nucleon inside the nucleus (with atomic mass number
$A$ and charge number $Z$) at an impact parameter $b$,
\begin{eqnarray}
    f_{a/A}(x,Q^2,b)&=& S_{a/A}(x,b)\left[ \frac{Z}{A}f_{a/p}(x,Q^2) \right.
    \nonumber \\
    & & +\left. (1-\frac{Z}{A}) f_{a/n}(x,Q^2)\right], \label{eq:shd}
\end{eqnarray}
is assumed to be factorizable into the parton distribution in a
nucleon $f_{a/N}(x,Q^2)$ and the nuclear modification factor
$S_{a/A}(x,b)$ which we take the parameterization used in HIJING
\cite{hijing} for now. The initial parton transverse momentum distribution
inside a projectile nucleon going through the target nucleon at an
impact parameter $b$ is then,
\begin{equation}
    g_A(k_T,Q^2)=\frac{1}{\pi \langle k^2_T\rangle_A} 
    e^{-k^2_T/\langle k^2_T\rangle_A}, \label{eq:ga}
\end{equation}
with a broadened variance
\begin{equation}
\langle k^2_T\rangle_A(Q^2)=\langle k^2_T\rangle_N(Q^2)
    +\delta^2(Q^2)(\nu_A(b) -1).
\end{equation}
The broadening is assumed to be proportional to the number of
scattering $\nu_A(b)$ the projectile suffers inside the nucleus.
For the purpose of considering the impact-parameter dependence of the $k_T$
broadening,  we will simply assume a hard sphere nuclear
distribution. Therefore, 
\begin{equation}
    \nu_A(b)=\sigma_{NN} t_A(b)=
    \sigma_{NN} \frac{3 A}{2\pi R_A^2}\sqrt{1-b^2/R_A^2},
\end{equation}
where $R_A=1.12 A^{1/3}$ fm and $\sigma_{NN}$ is the inelastic
nucleon-nucleon cross section.

We also assume that $k_T$ broadening during each nucleon-nucleon collision
$\delta^2$ also depends on the hard momentum scale $Q=P^{\rm jet}_T$, the
transverse momentum of the produced parton jet. Such a dependence is easy to
understand by comparing the role of $P^{\rm jet}_T$ in the multiple parton
scattering case to that in the initial and final state radiation associated
with a hard jet production. For example, unless the multiple scatterings are
far separated in space-time such that the propagating parton becomes on-shell,
the interaction cross of the propagating parton will depend on its
virtuality which in return depends on the scale of the hard scattering.
In Ref.~\cite{guo}, the $k_T$ broadening in a multiple parton
scattering scenario is related to the ratio of the leading-twist parton
distributions and higher-twist multiple parton distributions which apparently
depends on the momentum scale of the hard parton scattering.  
In this paper, we will use the
following scale-dependent $k_T$ broadening per nucleon-nucleon collision,
\begin{equation}
\delta^2(Q^2)=0.225\frac{\ln^2(Q/{\rm GeV})}{1+\ln(Q/{\rm GeV})}
 \;\;\;{\rm GeV^2}/c^2.
\end{equation}
Such a functional form is chosen to best fit the experimental data in $p+A$
collisions.
For $Q=2\sim 3$ GeV, $\delta^2=0.064 \sim 0.129$ GeV$^2/c^2$, which is
consistent with the value obtained from the analysis of $k_T$
broadening for $J/\Psi$ production in $p+A$ collisions \cite{gg,dns}.
We should point out that this value can be different from what one
gets from the analysis of $k_T$ broadening of Dell-Yan lepton pairs in
$p+A$ collisions, where mainly quarks and anti-quarks are involved in
the hard processes and there is no collision effect after the Dell-Yan
production point.

\begin{figure}
\centerline{\psfig{figure=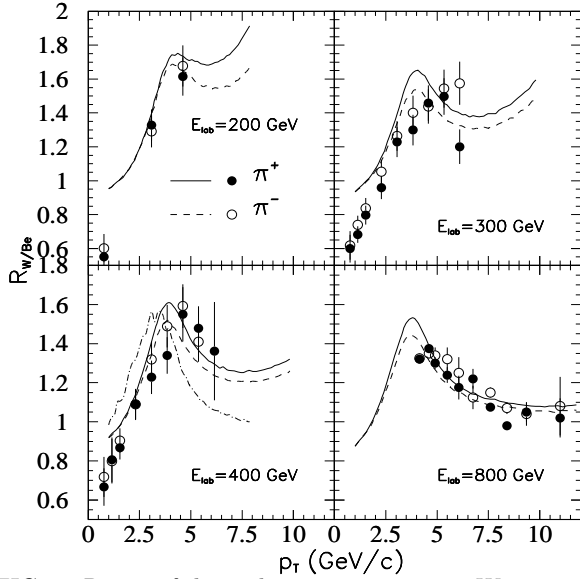,width=3.0 in,height=3.0in}}
\caption{ Ratios of charged pions spectra in $p+W$ over $p+Be$ each
  normalized by the atomic number of the target nucleus. The lines are the
  parton model calculation with $k_T$ broadening due to multiple parton
  scattering. The dot-dashed line in the third panel is the calculated result
  with a constant $k_T$ broadening $\delta^2=0.1$ GeV$^2$. Experimental data
  are from Refs.~\protect\cite{cronin-ex1,cronin-ex2}.}
\label{figsps9}
\end{figure}

    Using the above assumptions, we calculate the large $p_T$
particle spectra in $p+A$ collisions and compare to the spectra in
$p+p$ collisions. As we show in the last Section, particle spectra has
a strong isospin dependence in the parton model. In order to minimize this
known isospin dependence of the spectra in our study of nuclear
dependence of the spectra, we compare the spectra of $p+A$ for heavy
target with that for a very light nuclear target. Shown in
Fig.~\ref{figsps9} are our calculated ratios of charged pion spectra
(solid lines for $\pi^+$ and dot-dashed lines for $\pi^-$) in $p+W$ over
that of $p+Be$ each normalized by the atomic number of the target
nucleus. The experimental data are from Ref.~\cite{cronin-ex1,cronin-ex2}.
If there was no nuclear dependence due to multiple scattering, the ratios
would have a flat value of 1, modulo the residue isospin dependence
because of the small isospin asymmetry of the target nucleus. The
difference between our calculated ratios for $\pi^+$ and $\pi^-$ gives
the order of this isospin asymmetry effect. One can get rid of this
effect by using the ratios of $\pi^+ + \pi^-$ spectra.  As shown
in the figure, our model can roughly describe the general feature of
the nuclear dependence of the spectra at large $p_T$ due to multiple
parton scattering. The ratios should become smaller than 1 at very small
$p_T$ because of the absorptive processes as we
have mentioned. But here our perturbative calculation will eventually
breaks down because of the small momentum scale. At larger $p_T$, the
spectra are enhanced because of multiple parton scattering. As $p_T$
increases further, the ratios decrease again and saturate at about
1. The decrease follows the form of $1/p_T^2$ consistent with the
general features of high twist processes. Since the transverse
momentum broadening due to multiple parton scattering is finite, its
effect will eventually become smaller and disappear. Therefore, the
$p_T$ location of the maximum enhancement can give us the scale of
average transverse momentum broadening. We also show in the third panel as
dot-dashed line our calculation with a constant $k_T$ broadening,
$\delta^2=0.1$ GeV$^2$. While the result has the general feature of the
experimental data, the fit is not as good as the scale-dependent $k_T$
broadening.

\begin{figure}
\centerline{\psfig{figure=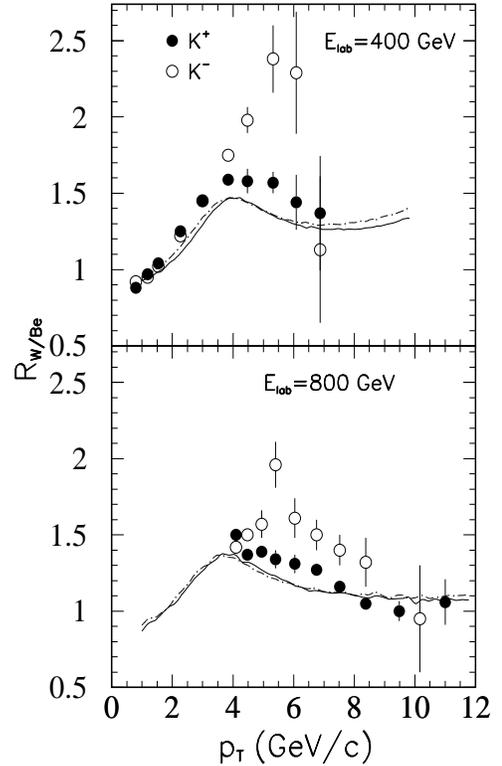,width=2.5 in,height=4.0in}}
\caption{ Ratios of charged kaons spectra in $p+W$ over $p+Be$ each
  normalized by the atomic number of the target nucleus. The solid
  (dot-dashed)lines are the parton model calculation for $K^+$ ($K^-$) with
  $k_T$ broadening due to multiple parton scattering. Experimental data are
  from Refs.~\protect\cite{cronin-ex1,cronin-ex2}.} 
\label{figsps10}
\end{figure}

    At low energies, $E_{\rm lab}=200$ GeV for example, the full structure
of multiple scattering cannot be revealed because of the finite phase space
constrained by the kinetic limit. Furthermore, the ratio will increase
at large $x_T=2p_T/\sqrt{s}$ because of the
so-called EMC effect \cite{emc} on nuclear structure function at large
$x$ which is incorporated into the parameterization of the nuclear
modification factor for parton distributions inside a nucleus. One
word of caution should also be
given here. Our calculation might not be reliable anymore near the
boundary of the available phase space ($p_T\sim\sqrt{s}/2$) where
overall energy and momentum conservation is very important which is not
imposed in QCD parton model.

We have also calculated the kaon spectra in $pA$ collisions. Show in
Fig.~\ref{figsps10} are the ratios of the kaon spectra in $p+W$ over $p+Be$
collisions (normalized by the atomic number). The agreement with the
experimental data is quite good for $K^+$. However, for $K^-$ there is quite
sizable discrepancy between the calculation and the data. This is might be
due to the flavor dependence of the $k_T$ broadening suffered by the
quarks. It is quite possible that $\bar u$ quarks have larger scattering
cross sections as they propagate through the nucleus and 
therefore experience larger $k_T$ broadening than $\bar s$ quarks. Such
effects will result in the observed difference in nuclear enhancement
between $K^-$ and $K^+$.

\section{$A+A$ Collisions: Jet quenching ?}

    It is straightforward to incorporate the initial $k_T$ broadening due to
multiple parton scattering in $A+A$ collisions. Since partons from
projectile and target beam both suffer 
multiple scattering before the hard process, their initial transverse
momentum distributions are broadened as given by Eq.~(\ref{eq:ga}). In
addition, the produced parton jets have to propagate through a
dense medium and interact with other produced particles in the
medium. Theoretical studies \cite{GW1,BDPS,BDMPS} show that a
fast parton propagating inside a dense partonic medium will suffer 
considerable amount of energy loss. Since
large $p_T$ particles are produced through the fragmentation of parton
jets with large transverse energy, parton energy loss will
definitely lead to the suppression of large $p_T$ particles. One can
describe this suppression phenomenologically via effective
fragmentation functions which are modified from their original forms
in the vacuum \cite{WH,WHS,wang98}. In this approach one can then
calculate the single inclusive particle spectra at large $p_T$ in
$A+B$ collisions as,
\begin{eqnarray}
  \frac{d\sigma^h_{AB}}{dyd^2p_T}=&K&\sum_{abcd} \int d^2b d^2r  
   t_A(r)t_B(|{\bf b}-{\bf r}|) \nonumber \\
   &\times& \int dx_a dx_b d^2k_{aT} d^2k_{bT} \nonumber \\
   &\times& g_A(k_{aT},Q^2,r) g_B(k_{bT},Q^2,|{\bf b}-{\bf r}|)  \nonumber \\ 
   &\times& f_{a/A}(x_a,Q^2,r)f_{b/B}(x_b,Q^2,|{\bf b}-{\bf r}|) \nonumber \\
    &\times& \frac{D_{h/c}(z_c,Q^2,\Delta L)}{\pi z_c}
  \frac{d\sigma}{d\hat{t}}(ab\rightarrow cd), \label{eq:nch_AA}
\end{eqnarray}
where $D_{h/c}(z_c,Q^2,\Delta L)$ is the modified effective
fragmentation function for produced parton $c$ which has to travel an
average distance $\Delta L$ inside a dense medium. We will not
elaborate on the modeling of the modified fragmentation functions
\cite{WH,WHS,wang98} here, except pointing out that it depends on two
parameters: the energy loss per scattering $\epsilon_c$ and the
mean free path $\lambda_c$ for a propagating parton $c$. The energy
loss per unit length of distance is then
$dE_c/dx=\epsilon_c/\lambda_c$. We also assume that a gluon's
mean free path is half of a quark and then the energy loss $dE/dx$ is
twice that of a quark. In principle, the energy loss $dE/dx$ should
depend on local parton density or temperature, the parton's initial
energy, the total distance and whether there is expansion
\cite{baier98}. These possible features will influence the final hadron
spectra and their phenomenological consequences have been studied in
detail in a previous publication \cite{wang98}. In this paper, we will
simply assume a constant energy loss and study the average effect of
parton energy loss in dense matter.

After integration over the impact parameter space, Eq.~(\ref{eq:nch_AA})
will give us the inclusive hadron spectra for minimum-biased events of
$A+B$ collisions. If one neglects nuclear effects (parton shadowing,
$k_T$ broadening and jet quenching), the resultant spectra should be
exactly proportional to $AB$ which is the averaged number of binary
nucleon-nucleon collisions. Very often experimentalists also measure
inclusive hadron spectra for certain classes of events with different
centrality cuts (total transverse energy $E_T$ or charged multiplicity
in the central region). Since one cannot directly measure the impact
parameters in each class of centrality, a theoretical model of
correlation between impact parameter $b$ and the total transverse
energy $E_T$ (or charged multiplicity) has to be introduced in order
to calculate the inclusive cross section for events with different
centrality cuts. In this paper, we will use the correlation function
$P_{AB}(E_T,b)$ [normalized to $\int dE_T P_{AB}(E_T,b)=1$] introduced
in Ref.~\cite{kns2} which assumes a wounded nucleon model for
transverse energy production.  The differential cross section
\begin{equation}
    \frac{d\sigma_{AB}}{dE_T}=\int d^2b
    [1-e^{-\sigma_{NN}T_AB(b)}]P_{AB}(E_T,b) \label{eq:dsdet}
\end{equation}
has been shown to reproduce the experimental data of NA35
and NA49 \cite{na49} very well, where 
$T_{AB}(b)=\int d^2r t_A({\bf r}) t_B({\bf b}-{\bf r}) $ is the overlap
function for $A+B$ collisions at impact parameter $b$. We refer
readers to Ref.~\cite{kns2} for the details of this model.

    After incorporating the correlation function between impact
parameter and transverse energy in Eq.~(\ref{eq:nch_AA}), the inclusive
cross section of large $p_T$ hadron production in events with
centrality cut $E_T \in (E_T^{\rm min},E_T^{\rm max})$ can be shown to be
roughly proportional to
\begin{equation}
\langle N_{\rm binary}\rangle \equiv 
\int d^2b \int_{E_T^{\rm min}}^{E_T^{\rm max}} dE_T P_{AB}(E_T,b)
T_{AB}(b) \label{eq:binary}
\end{equation}
which is just the average number of binary nucleon-nucleon
collisions. For minimum-biased events, this number is just $AB$. Since
the actual $d\sigma_{AB}/dE_T$ distribution depends on each experiment's
coverage of phase space and the detector's calibration, we will
choose the values of $E_T^{\rm min}$ and $E_T^{\rm max}$ so that the
fraction of the integrated cross section in Eq.~(\ref{eq:dsdet}) within
the $E_T$ range matches the experimental value of a given centrality cut.

\begin{figure}
\centerline{\psfig{figure=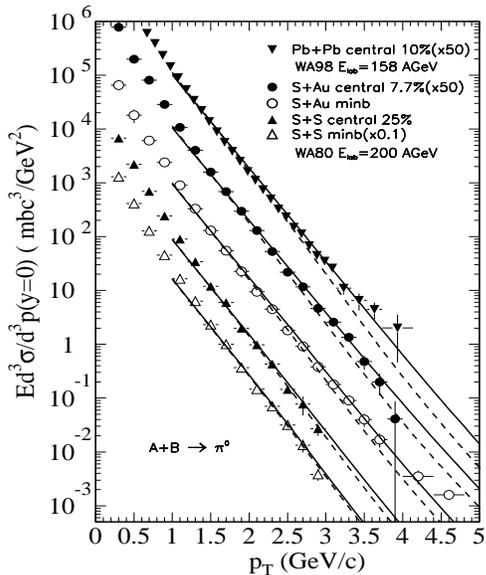,width=2.5 in,height=3.0in}}
\caption{ Single-inclusive spectra of $\pi^0$ in $S+S$, $S+Au$ and $Pb+Pb$
  collisions (both minimum-biased and central events) at the CERN SPS
  energies. The solid lines are QCD parton model calculations with $k_T$
  broadening due to initial multiple parton scattering and the dashed lines
  are without.  Experimental data are from Refs.~\protect\cite{wa80,wa98}.}
\label{figsps11}
\end{figure}

Shown in Fig.~\ref{figsps11} are the calculated inclusive spectra for
produced $\pi_0$ in $S+S$, $S+Au$ and $Pb+Pb$ collisions, both
minimum-biased and central events. The QCD parton model calculations
with the $k_T$ broadening due to initial multiple scattering (solid lines) 
agree with the experimental data (WA80 and WA98) \cite{wa80,wa98} well
very at $p_T$ above 1 GeV/$c$. No parton energy loss has been assumed
in the calculations. The dashed lines are the spectra in
$pp$ collisions at the same energy multiplied by the nuclear
geometrical factor as given in Eq.~(\ref{eq:binary}).  The difference
between the solid and dashed lines is simply caused by effects
of $k_T$ broadening and nuclear modification of parton distributions
inside nuclei. These effects are similar as in $p+A$ collisions and
are more important in collisions of heavier nuclei. Without these
nuclear effects the high $p_T$ hadron spectra in $A+B$ collisions are
exactly proportional to the average number of binary collisions as
shown by the dashed lines. This is a common characteristic of hard
processes in $p+A$ and $A+B$ collisions. Because of absorptive
processes, low $p_T$ particle production, which can be considered as
coherent over the dimension of nuclear size, has much weaker
$A$-dependence. In the wounded-nucleon model, soft particle production
cross section is proportional to the average number of wounded nucleons
which is much smaller than the number of binary collisions. By
studying the transition of the scaling property of the hadron spectra
from low to high $p_T$ values, one can then determine at what $p_T$ value
the underlying mechanism of hadron production become dominated by hard
processes. This will be the subject of another study.

\begin{figure}
\centerline{\psfig{figure=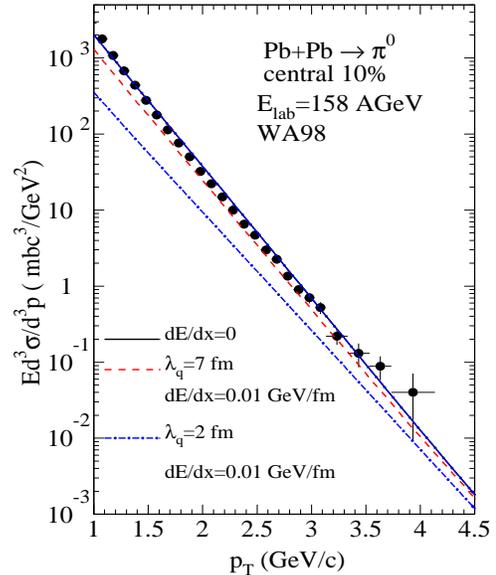,width=2.5 in,height=3.0in}}
\caption{Parton model calculations of single-inclusive spectra of $\pi^0$
  with different values of parton energy loss $dE_q/dx$ and mean free path
  $\lambda_q$ in central $Pb+Pb$ collisions at $E_{\rm lab}=158$ AGeV/. The
  Experimental data are from Ref.~\protect\cite{wa98}.}
\label{figsps12}
\end{figure}

In the parton model calculations presented so far, effects of parton energy
loss have not been considered yet. If there is parton energy loss and the
radiated gluons become incoherent from the leading parton, the resultant
leading hadron spectra at large $p_T$ from the parton fragmentation should
be suppressed as compared to $p+p$ and $p+A$ collisions. As we have shown,
however, the parton model calculations without parton energy loss fit the
experimental data very well. To find out the consequences of an effective
parton energy loss in the hadron spectra and how the experimental data
constrain its value at the CERN SPS energy, we show in Fig.~\ref{figsps12}
as the dot-dashed line 
the calculated $\pi_0$ spectra in central $Pb+Pb$ collisions with
$dE/dx=0.01 $ GeV/fm.  In the calculation (dot-dashed line) we have assumed
the energy loss per scattering $\epsilon_q=0.02$ GeV and the mean free path
$\lambda_q=2$ fm ($dE_q/dx=\epsilon_q/\lambda_q$). In our
model~\cite{WH,WHS}, the modification of the fragmentation function is
sensitive to both parameters. The mean free path is a measure of the strength
of the interaction between the leading parton and medium while energy loss
$dE/dx$ reflects the degree of attenuation induced by the interaction. Shown
as the dashed line in Fig.~\ref{figsps12} is a calculation with the same
$dE_q/dx=0.01$ GeV/fm but with a mean free path $\lambda_q=7$ fm which is
about the average total length a parton will travel through in a cylindrical
system with a radius of a $Pb$ nucleus. Such a scenario of weak interaction
and small energy loss is barely consistent with the systematics of the
experimental data. It is clear that the large $p_T$ spectra in $Pb+Pb$
collisions at the CERN SPS energy can put very stringent limits on the
interaction of energetic partons with dense medium and the induced energy
loss. Within the parton model, one can exclude from the observed hadron
spectra a parton energy loss larger than $dE_q/dx=0.01$ GeV/fm and a mean
free path shorter than $\lambda_q=7$ fm. This is much smaller than the most
conservative estimate of parton energy loss in a dense medium
\cite{GW1,BDPS,BDMPS}. According to the most conservative estimate in 
Ref. \cite{BDMPS}, a quark should still have an energy loss 
$dE/dx\approx 0.2$ GeV/fm in a cold nuclear matter of 10 fm in transverse
size. Our constraint is 20 times smaller than this conservative limit, even
though the dense matter in $AA$ collisions should at least consists of hot
hadronic matter which is much denser than the normal cold nuclear matter.

There are several implications one can draw from this analysis. Most of the
recent theoretical estimates of parton energy loss are based on a scenario of
a static and infinitely large dense parton gas. If the system produced in a
central 
$Pb+Pb$ collision only exists for a period of time shorter than the
interaction mean free path of the propagating parton, one then should not
expect to see any significant parton energy loss. Using the measured
transverse energy production $dE_T/d\eta\approx$ 405 GeV \cite{na49} and
a Bjorken scaling picture, one can indeed estimate \cite{wang98-2} that the
life time of the dense system in central $Pb+Pb$ collisions is only about 
2 -- 3 fm/$c$ before the density drops below a critical value of
$\epsilon_c\approx 1$ GeV/fm$^3$. Even if we assume that a dense partonic
system is formed in central $Pb+Pb$ collisions, this optimistic estimate
of the life time of the system could still be smaller than the mean free
path of the 
propagating parton inside the medium. Thus, one does not have to expect a
significant effect of parton energy loss on the final hadron spectra at
large $p_T$. Otherwise, it will be difficult to reconcile the absence of
parton energy loss with the strong parton interaction which maintains a
long-lived partonic system. 

Another conclusion one can also make is that the dense hadronic matter which
has existed for a period of time in the final stage of heavy-ion collisions
does not cause any apparent parton energy loss or jet quenching. A high
$p_T$ physical pion from jet fragmentation has a very long formation
time. One does not have to worry about its scattering with other soft
hadrons in the system which could cause suppression of high $p_T$ pion
spectra. We still do not understand the reason why a fragmenting parton does
not loss much energy when it propagates through a dense hadronic
matter. However, it might be related to the absence of energy loss to the
quarks and anti-quarks prior to Drell-Yan hard processes in $p+A$ and $A+A$
collisions. This observation will make jet quenching a better probe of a
long-lived partonic matter since one does not have to worry about the
complications arising from the hadronic phase of the evolution. If one
observes a dramatic suppression of high $p_T$ hadron spectra at the BNL RHIC
energy as predicted \cite{WG92,WH,WHS,wang98}, then it will clearly indicate
an initial condition very different from what has been reached at the CERN
SPS energy.

\section{Predictions at RHIC: Importance of $p+A$ Experiments}

\begin{figure}
\centerline{\psfig{figure=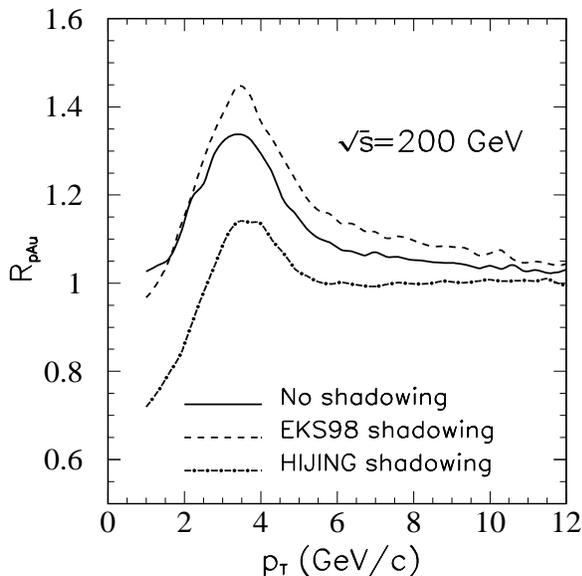,width=3.0 in,height=3.0in}}
\caption{ Predictions for the ratio of single charged hadron spectra in
  $p+Au$ over $p+p$ collisions normalized by the average number of binary
  collisions (or $A$ ) at $\sqrt{s}=200$ GeV. Different lines are for
  different parameterizations of shadowing or nuclear modification of parton
  distributions.}
\label{figsps13}
\end{figure}

As we have seen in the calculation of large $p_T$ particle spectra in
$p+p (\bar{p})$ collisions, the initial $k_T$ smearing is most important at
low energies where the original particle spectra from jet production falls
off very rapidly with $p_T$. The effect of $k_T$ broadening due to 
multiple parton scattering in $p+A$ collisions follows the same line.
At higher energies ($\sqrt{s}> 50 $ GeV), when the original jet
spectra become much flatter, the effects of both the initial or
intrinsic $k_T$ and nuclear $k_T$ broadening will become smaller, as we can
clearly see in both our calculation and the experimental data in
Figs.~\ref{figsps8} and \ref{figsps5}. At the BNL RHIC collider energy, the
nuclear enhancement of large $p_T$ spectra will then become smaller.
Shown in Fig.~\ref{figsps13} are the calculated ratios of charged hadron
spectra in $p+Au$ over that in $p+p$ normalized to the averaged number of
binary nucleon collisions,
\begin{equation}
  R_{AB}(p_T)\equiv \frac{d\sigma^h_{AB}/dyd^2p_T}
  {\langle N_{\rm binary}\rangle d\sigma^h_{pp}/dyd^2p_T} \label{eq:ratio}
\end{equation}
where $\langle N_{\rm binary}\rangle=A$ for minimum-biased events of $p+A$
collisions. We can see that the enhancement due to multiple parton
scattering will not disappear at the RHIC energy. There is still
about 20--50\% enhancement at moderately large $p_T$ around 4 GeV/$c$. 
It then disappears very quickly at larger $p_T$. 

At the BNL RHIC energy, $\sqrt{s}=200$ GeV, nuclear modification of the
gluon distribution will become important for hadron spectra at moderately large
$p_T$. Such modifications for quark distributions have been measured in
deeply inelastic lepton-nucleus collisions \cite{eks}. However,
the nuclear effects on the gluon distribution have not been directly
measured. In the calculation shown as the dashed line in
Fig.~\ref{figsps13} we have used a recent parameterization of the nuclear
modification factors $S_{a/A}(x,Q^2)$ [Eq.~(\ref{eq:shd})] by Eskola, Kolhinen
and Salgado \cite{eks} which is based on global fits to the most recent
collection of data available and some model on nuclear modification of gluon
distribution. This result is quite different from the calculation using
HIJING \cite{hijing} parameterization. In EKS98 parameterization, QCD
evolution equation has been used to take into account the $Q^2$ scale
dependence of the nuclear modification which is absent in HIJING
parameterization. Because of QCD evolution, the nuclear shadowing effects at
small $x$ become smaller with increasing $Q^2$. This is why the hadron
spectrum in $p+A$ collisions using EKS parameterization is larger than that
using HIJING 
parameterization. The EKS98 parameterization also has a gluon anti-shadowing
which is larger than any parameterizations before. Hadron spectra in
a moderately large $p_T$ range at the BNL RHIC energy mainly come from
fragmentation of gluon jets. This is why the EKS98 result in
Fig.~\ref{figsps13} is larger than that without nuclear 
modification of parton distributions (solid line) in the $p_T$ range where
anti-shadowing becomes relevant. In addition, the impact
parameter dependence of the nuclear modification of parton distributions is
not implemented in the calculation using EKS98 parameterization.
Though EKS98 parameterization has taken into account information about
nuclear shadowing of gluon distribution from the measured scale evolution of
the structure functions of different nuclei, such a procedure is also model
dependent. The result will depend on whether one includes the higher
gluon-density terms in the QCD evolution equation.
Before a direct measurement of the nuclear effect on gluon distribution
inside a nucleus, one should consider it one of the uncertainties in
predicting the hadron spectra in $p+A$ collisions at the BNL RHIC energy.

As pointed out in Sec. II, the shape of the nuclear modification of hadron
spectra at large $p_T$ is a result of multiple parton scattering inside a
nucleus and their corresponding absorptive corrections. The shape is
remarkably similar to the nuclear modification of parton distributions. In
fact, nuclear modification of parton distributions in small $x$ region
(shadowing and anti-shadowing) can be explained in a multiple scattering
model \cite{lu}. Therefore, one could have double counted the same effect if
he includes both the Cronin effect (or $k_T$ broadening) and nuclear
modification of the parton distributions in the calculation of hadron
spectra in $p+A$ collisions. Given these uncertainties, it is therefore
extremely important to have a systematic study of $p+A$ collisions at the
BNL RHIC energy. Such an effort is pivotal to unravel any other effects
in the hadron spectra caused by the formation of dense partonic matter in
$A+A$ collisions.

\begin{figure}
\centerline{\psfig{figure=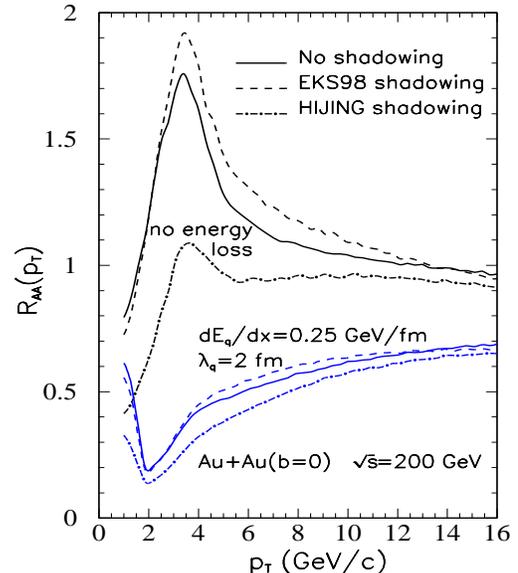,width=2.6 in,height=3.0in}}
\caption{ Predictions for the ratio of single charged hadron spectra in
  central $Au+Au$ over $p+p$ collisions normalized by the average number of
  binary collisions at $\sqrt{s}=200$ GeV. Different lines are for
  different parameterizations of shadowing or nuclear modification of parton
  distributions. The upper set of lines are without parton
  energy loss and lower set are with parton energy loss $dE_q/dx=0.25$
  GeV/fm and mean free path $\lambda_q=2$ fm.}
\label{figsps14}
\end{figure}

Shown in Fig.~\ref{figsps14} are predictions for the ratio of hadron
spectra in central $Au+Au$ collisions over that in $p+p$ [as defined in
Eq.~(\ref{eq:ratio})], with and without parton energy loss. In both
cases, uncertainties in the nuclear modification of parton distributions are
still important. If there is significant parton energy loss in the dense
medium at the BNL RHIC energy,  leading hadrons from jets which are
produced in the center of a overlapped region will be suppressed. As we can
see this is where
the Cronin effect and the effect of nuclear modification of parton
distributions are the strongest. This is why the relative uncertainty caused
by nuclear parton distributions is reduced compared to the case when there
is no parton energy loss. In the calculations, we also included possible
contribution to hadron production from soft processes at low $p_T$ as
implemented in 
Ref.~\cite{wang98}. When hadrons from hard parton scattering are suppressed,
these soft contributions might become dominant at intermediate $p_T$. This
is why the spectra ratio increases again when $p_T<2$ GeV/$c$ in
Fig.~\ref{figsps14}. The spectra at lower $p_T$ could also be broadened
due to the rescattering effect which also drives the system into local
equilibrium. We should emphasize that the soft particle spectra we use here
are extremely schematic and qualitative. This is another uncertainty one
should keep in mind which could affect the shape of spectra at intermediate
$p_T$. Nevertheless,the spectra at moderately large $p_T$ is most sensitive
to parton energy loss in dense  medium. Therefore, one should study hadron
spectra in $p+A$ very carefully and pin 
down the uncertainty due to multiple scattering and parton (anti)shadowing
as much as possible. Only then one can draw more accurate conclusions
about parton energy loss from the single inclusive hadron spectra in
heavy-ion collisions at the BNL RHIC energy.

\section{Conclusions and Discussions}

In this paper, we have analyzed systematically large $p_T$ hadron
spectra in $p+p$, $p+A$ and $A+A$ collisions from CERN SPS to BNL RHIC
energies within a QCD parton model. We found that both the initial $k_T$
in $p+p$ collisions and the $k_T$ broadening due to multiple parton
scattering in $p+A$ collisions are important to describe the experimental
data within the parton model calculations. The value of initial $k_T$ in
order to fit the data is found to be larger than the conventional
value of $300-500$ MeV for intrinsic $k_T$ according to the uncertainty
principle. This finding is also 
consistent with analysis of Drell-Yan data \cite{field} and recent study
\cite{photon} of direct photon and pion production at around CERN SPS energy
range. In both studies, one found that an intrinsic $k_T$ of the order of 1
GeV/$c$ is needed to describe the data within NLO parton model
calculations. Since we only used LO pQCD calculation, we have to introduce
some $Q^2$ dependence of the initial $k_T$ induced by initial-state
radiation processes in order to fit the experimental data at low collider
energies. This parton model with phenomenologically included intrinsic $k_T$
describes very well the energy and isospin dependence of the hadron spectra. 

The parton model calculation can also describe the large $p_T$ pion spectra
in heavy-ion collisions at the CERN SPS energies very well, both the $A$ or
centrality and energy dependence. There is no evidence of proposed parton
energy loss caused by dense partonic matter. Based on recent theoretical
estimates \cite{GW1,BDPS,BDMPS} of parton energy loss in dense partonic
matter, one should expect a parton energy loss in the order of $dE/dx \sim 2
- 4$ GeV/fm. The absence of such energy loss in large $p_T$ hadron spectra
implies that either there is no such dense partonic matter formed or the
life time of such medium is smaller then the mean free path of the parton
interaction inside such a medium. It also tells us that the hadronic matter
which must have existed for a period of time in heavy-ion collisions at the
CERN SPS will not cause apparent energy loss or jet quenching
effect. Therefore, if one observes suppression of high $p_T$ hadrons at the
BNL RHIC energy, it will unambiguously reflect an initial condition very
different from what has been achieved at the CERN SPS.

As pointed out in a recent paper by Gyulassy and Levai \cite{MGPL}, even
though HIJING \cite{hijing} Monte Carlo model fails to reproduce the hadron
spectra in $p+p$, $p+A$ and $S+S$ (and $S+Au$) data at the CERN SPS energy,
it accidentally reproduces the central $Pb+Pb$ data very well because some
unique form of transverse momentum kick introduced in HIJING to the
end-points of strings each time they suffer an interaction. A hydrodynamic
model calculation \cite{dumitru} with significant transverse expansion
can also be used to fit the high $p_T$ hadron spectra. So one 
should wonder how one can make sure that the high $p_T$ hadron production
in $A+B$ collisions is indeed dominated by hard parton scattering. As we
have proposed earlier \cite{wang98-2,wang92}, measurement of
two-particle correlation in azimuthal angle in the transverse plane should
be able to distinguish these models from parton model. In the parton model,
jets are always produced in pairs and back-to-back in the transverse
plane. High $p_T$ particles from jet fragmentation should then have strong
back-to-back correlation. Neither hydro-dynamical model nor multiple
scattering of string end-points can give such correlation. If such
correlation is seen, one can then study the $p_T$ dependence of the
correlation to find out at what $p_T$ value the correlation disappears. One
can then at least quantify above what $p_T$ value thermal-hydro model
can be ruled out as the underlying particle production mechanism. Such a
study is underway and will be reported elsewhere.

At the BNL RHIC energy, things will become a little cleaner. As pointed out by
Gyulassy and Levai \cite{MGPL}, most of hadron production at moderate $p_T$
will be dominated by gluon production and are not influenced by the fate of
strings or valence quarks (except net baryons). The spectra should be more
sensitive to parton energy loss. Moreover, the hadron spectra in $p+A$ and
$A+A$ collisions without parton energy loss will have a unique shape similar
to those shown in Fig.~\ref{figsps13} at high energies which can not be
fully revealed at the CERN SPS energies. This unique shape will be hard to
explain by any thermal-hydrodynamic models. However, as we have demonstrated
in the previous section, there are still some uncertainties related to
nuclear modification of parton distributions. It is therefore very important
to have a systematic study of $p+p$, $p+A$ and $A+A$ collisions at the BNL
RHIC energy in order to make more quantitative conclusions about parton
energy loss in dense medium from the single hadron spectra at moderately
high $p_T$.

\section*{Acknowledgements}
This work was supported by the Director, Office of Energy
Research, Office of High Energy and Nuclear Physics, Divisions of 
Nuclear Physics, of the U.S. Department of Energy under Contract No.\
DE-AC03-76SF00098 and DE-FG03-93ER40792. The author wish to thank
K. J. Eskola for comments and for providing the EKS98 parameterization of
the nuclear modification of parton distributions.

\end{multicols}  
\end{document}